\documentclass[aps,prl,twocolumn,superscriptaddress]{revtex4}

\usepackage[dvips]{graphicx}
\usepackage{color}
\usepackage{bm}
\usepackage{amssymb}
\renewcommand{\vec}[1]{\boldsymbol{#1}}
\usepackage[utf8]{inputenc}
\usepackage{times}

\begin{document}

\title{Structural, electronic, and magnetic properties of nearly-ideal $J_{\rm eff}$~$=$~1/2 iridium halides}

\author{D. Reig-i-Plessis}
\affiliation{Department of Physics and Materials Research Laboratory, University of Illinois at Urbana-Champaign, Urbana, IL 61801, USA}
\affiliation{Department of Physics and Astronomy and Quantum Matter Institute, University of British Columbia, Vancouver, BC V6T 1Z1, Canada}

\author{T.A. Johnson}
\affiliation{Department of Physics and Materials Research Laboratory, University of Illinois at Urbana-Champaign, Urbana, IL 61801, USA}

\author{K. Lu}
\affiliation{Department of Physics and Materials Research Laboratory, University of Illinois at Urbana-Champaign, Urbana, IL 61801, USA}

\author{Q. Chen}
\affiliation{Department of Physics and Astronomy, University of Tennessee, Knoxville, TN 37996, USA}

\author{J.P.C. Ruff}
\affiliation{Cornell High Energy Synchrotron Source, Cornell University, Ithaca, NY 14853, USA}

\author{M.H. Upton}
\affiliation{Advanced Photon Source, Argonne National Laboratory, Lemont, IL 60439, USA}

\author{T.J. Williams}
\affiliation{Neutron Scattering Division, Oak Ridge National Laboratory, Oak Ridge, TN 37831, USA}

\author{S. Calder}
\affiliation{Neutron Scattering Division, Oak Ridge National Laboratory, Oak Ridge, TN 37831, USA}

\author{H.D. Zhou}
\affiliation{Department of Physics and Astronomy, University of Tennessee, Knoxville, TN 37996, USA}

\author{J.P. Clancy}
\affiliation{Department of Physics and Astronomy, McMaster University, Hamilton, ON L8S 4M1, Canada}

\author{A.A. Aczel}
\email{aczelaa@ornl.gov}
\affiliation{Department of Physics and Astronomy, University of Tennessee, Knoxville, TN 37996, USA}
\affiliation{Neutron Scattering Division, Oak Ridge National Laboratory, Oak Ridge, TN 37831, USA}

\author{G.J. MacDougall}
\email{gmacdoug@illinois.edu}
\affiliation{Department of Physics and Materials Research Laboratory, University of Illinois at Urbana-Champaign, Urbana, IL 61801, USA}

\date{\today}

\begin{abstract}
Heavy transition metal magnets with $J_{\rm eff}$~$=$~1/2 electronic ground states have attracted recent interest due to their penchant for hosting new classes of quantum spin liquids and superconductors. Unfortunately, model systems with ideal $J_{\rm eff}$~$=$~1/2 states are scarce due to the importance of non-cubic local distortions in most candidate materials. In this work, we identify a family of iridium halide systems [i.e. K$_2$IrCl$_6$, K$_2$IrBr$_6$, (NH$_4$)$_2$IrCl$_6$, and Na$_2$IrCl$_6 \cdotp $ 6(H$_2$O)] with Ir$^{4+}$ electronic ground states in extremely close proximity to the ideal $J_{\rm eff}$~$=$~1/2 limit, despite a variation in the low-temperature global crystal structures. We also find ordered magnetic ground states for the three anhydrous systems, with single crystal neutron diffraction on K$_2$IrBr$_6$ revealing Type-I antiferromagnetism. This spin configuration is consistent with expectations for significant Kitaev exchange in a face-centered-cubic magnet.
\end{abstract}

\maketitle

\section{I. Introduction}

Heavy transition metal magnets with $d^5$ ions in an ideal octahedral local environment have been shown to host a $J_{\rm eff}$~$=$~1/2 spin-orbit-assisted Mott insulating state \cite{08_kim}. The initial interest in materials with this exotic electronic ground state was largely twofold. First, the canonical $J_{\rm eff}$~$=$~1/2 system Sr$_2$IrO$_4$ shares common phenomenology with cuprate superconductors \cite{14_kim, 16_kim} and has been predicted to superconduct upon doping \cite{11_wang, 13_watanabe}, although this remains to be verified by experiments. Second, a honeycomb lattice of $J_{\rm eff}$~$=$~1/2 moments provides an experimental platform for testing predictions of the Kitaev model \cite{06_kitaev, 09_jackeli}, which is exactly solvable and yields a `Kitaev spin liquid' ground state with Majorana fermion excitations. Several candidate systems, including the honeycomb iridates Na$_2$IrO$_3$ \cite{10_singh, 11_jiang}, $\alpha$-Li$_2$IrO$_3$ \cite{12_singh, 13_cao_2} and $\alpha$-RuCl$_3$ \cite{14_plumb, 15_johnson, 15_sears} have now been intensively investigated to look for evidence of this exotic state-of-matter. Although additional interactions beyond the bond-directional Kitaev term, such as direct Heisenberg exchange or off-diagonal exchange $\Gamma$, typically stabilize ground states with ordered spin configurations instead \cite{13_chaloupka, 14_rau}, it is now well-established that at least $\alpha$-RuCl$_3$ \cite{15_sandilands, 16_banerjee, 17_banerjee} is proximate to the desired spin liquid state.

Significantly fewer studies have investigated $J_{\rm eff}$~$=$~1/2 magnetism beyond Sr$_2$IrO$_4$ and the honeycomb lattice, although characterizing the metal-insulator transitions in pure and doped Sr$_3$Ir$_2$O$_7$ \cite{13_li, 14_delatorre, 14_dhital, 15_hogan, 16_ding, 16_donnerer, 18_donnerer} and searching for Kitaev materials with $J_{\rm eff}$~$=$~1/2 moments on other sublattices \cite{17_winter} are two enduring themes. Pioneering work by Kimchi and Vishwanath \cite{14_kimchi} provides strong motivation for the latter topic, as they have pointed out that the Kitaev interaction is symmetry-allowed on several sublattices beyond the honeycomb, including the triangular, Kagome, pyrochlore, hyperkagome, and face-centered-cubic (fcc) geometries. They have also used Luttinger-Tisza analysis to calculate the classical Heisenberg-Kitaev phase diagrams for these different cases, which are quite rich and contain several regimes with extensive degeneracy that they label `quantum phases'. Follow-up work determined the phase diagrams for the Heisenberg-Kitaev (or Heisenberg-Kitaev-$\Gamma$) models on the triangular and fcc lattices using other classical or quantum treatments, and these studies indicate that exotic vortex crystal, spin liquid, incommensurate spiral, or nematic states may be realized \cite{15_becker, 15_catuneanu, 15_li, 15_cook, 16_shinjo, 19_revelli}. Therefore, there is strong incentive to identify $J_{\rm eff}$~$=$~1/2 fcc or triangular lattice systems that can be used to test these predictions.

Ideal $J_{\rm eff}$~$=$~1/2 magnets have been difficult to find, and notably the local environments of the Ir$^{4+}$ ions are non-cubic in all known iridates based on corner-sharing, edge-sharing, or face-sharing IrO$_6$ octahedra. While it has been argued that many iridates are effectively in the $J_{\rm eff}$~$=$~1/2 limit \cite{13_gretarsson, 14_calder, 19_revelli, 19_aczel}, significant concerns remain that small non-cubic deviations of the electronic wave functions may mask the intrinsic properties of the spin-orbit-assisted insulating state, preclude superconductivity in doped $J_{\rm eff}$~$=$~1/2 magnets, prevent the realization of the Kitaev model in honeycomb systems, or complicate comparisons between theoretical and experimental phase diagrams. Making matters worse, there are no known $J_{\rm eff}$~$=$~1/2 fcc or triangular lattice iridates with the IrO$_6$ octahedral connectivity described above. Briol and Haule pointed out that both these limitations may be overcome by extending the search for $J_{\rm eff}$~$=$~1/2 magnets to materials with isolated Ir$X_6$ ($X$~$=$~anion) octahedra \cite{15_birol}, as spacing the magnetic ions further apart is expected to generate extremely narrow $d$-bands and hence promote robust insulating states. These ideas have now been explored in the triangular lattice iridate Ba$_3$IrTi$_2$O$_9$ \cite{15_becker} and the fcc (or quasi-fcc) Ir$^{4+}$ double perovskite iridates \cite{13_cao, 15_cook, 16_aczel, 19_aczel}, which feature isolated Ir$_2$O$_9$ bi-octahedra and IrO$_6$ octahedra respectively. While Ir/Ti site mixing leads to complications for Ba$_3$IrTi$_2$O$_9$ \cite{12_dey, 17_lee}, Ba$_2$CeIrO$_6$ remains cubic down to 4 K \cite{19_aczel} and therefore was initially expected to host an ideal $J_{\rm eff}$~$=$~1/2 insulating state. It was surprising then when resonant inelastic x-ray scattering (RIXS) measurements revealed a small splitting of the excited $J_{\rm eff}$~$=$~3/2 quartet into two doublets, which may arise from undetected local distortions at the Ir sites \cite{19_aczel, 19_revelli}.

A related topic of recent interest has been the prospect of realizing significant Kitaev interactions in materials like Ba$_3$IrTi$_2$O$_9$ \cite{15_becker} and the Ir$^{4+}$ double perovskite iridates, since some of the exotic phases predicted above for $J_{\rm eff}$~$=$~1/2 triangular and fcc magnets are experimentally observed. Initial work on the quasi-fcc double perovskites La$_2$MgIrO$_6$ and La$_2$ZnIrO$_6$ showed that the thermodynamic properties \cite{15_cook} and powder-averaged spin wave spectra \cite{16_aczel} can be explained by a magnetic Hamiltonian with dominant nearest neighbor (NN) antiferromagnetic (AFM) Kitaev exchange, although subsequent theoretical calculations on fcc Ba$_2$CeIrO$_6$ support a dominant NN AFM Heisenberg exchange with a smaller (but notably still significant) NN AFM Kitaev term. While the relative strength of the NN Heisenberg and Kitaev terms remains an open question, there is now a growing consensus that the latter is responsible for stabilizing the same Type-I (A-type) AFM order in several double perovskite iridates \cite{13_cao, 15_cook, 16_aczel, 19_aczel} and therefore cannot be neglected in any accurate microscopic model of these materials. These findings provide optimism that a significant portion of the Heisenberg-Kitaev phase diagram can be accessed in the laboratory by $J_{\rm eff}$~$=$~1/2 fcc magnets, constructed from isolated Ir$X_6$ octahedra.

To date, the family of Ir$^{4+}$ double perovskite iridates consists of only four well-studied members with a single magnetic sublattice: Ba$_2$CeIrO$_6$, Sr$_2$CeIrO$_6$, La$_2$MgIrO$_6$ and La$_2$ZnIrO$_6$ \cite{95_ramos, 96_battle, 99_harada, 00_harada, 00_wakeshima, 13_cao, 13_panda, 15_cook, 16_aczel, 16_ferreira, 16_karungo, 18_han, 18_iakovleva, 19_aczel}. Unfortunately, it is now well-established that none of these systems are ideal $J_{\rm eff}$~$=$~1/2 magnets, and large single crystal growth is not possible, severely limiting the impact of many advanced characterization studies. As an interesting set of alternatives, Birol and Haule \cite{15_birol} identified several other $J_{\rm eff}$~$=$~1/2 fcc magnet candidates of the form $A_2 B X_6$, where $A$ is often an alkali metal, $B$ is a 4$d$/5$d$ transition metal, and $X$ is an anion (typically a halogen ion), though they have received surprisingly little attention in this context. Antifluorites of this form were investigated by Wyckoff and Posnjak as early as the 1920s \cite{wyckoff_textbook}, and in the 1960s - 1980s were actively studied as model systems for exploring NN-NNN (next nearest neighbor) fcc magnetism, lattice dynamics, and structural phase transitions \cite{80_armstrong}. There is now a new impetus to investigate these materials using modern instrumentation to assess their $J_{\rm eff}$~$=$~1/2 candidacy and elucidate the role that Kitaev interactions may play in establishing their magnetic properties. One recent study has discussed these two issues for the particular case of K$_2$IrCl$_6$ \cite{19_khan}, but even for this material spectroscopic evidence for the $J_{\rm eff}$~$=$~1/2 state is still lacking.

The antifluorite structure can accommodate water molecules in several different ways \cite{18_bao}, and the anion $X$ can be tuned easily; this flexibility leads to many more $J_{\rm eff}$~$=$~1/2 magnet candidates as compared to the double perovskite structure. Sizable single crystals of several family members can also be grown by solution or chemical vapor transport methods. These advantages make the $A_2 B X_6$ family particularly attractive for systematic studies of heavy transition metal magnetism on the fcc lattice, which is the focus of the current work. In subsequent sections, we explore the evolution of the crystal structure, electronic properties, and magnetic properties when modifying the $A$ ion, the $X$ ion, or adding water molecules, by providing detailed characterization data on the materials K$_2$IrCl$_6$, K$_2$IrBr$_6$, (NH$_4$)$_2$IrCl$_6$, and Na$_2$IrCl$_6 \cdotp $ 6(H$_2$O). Diffraction techniques identify both fcc antifluorite and highly distorted structures in this family of materials, yet x-ray spectroscopy measurements reveal unprecedented proximity to $J_{\rm eff}$~$=$~1/2 electronic ground states for the Ir$^{4+}$ ions in {\it all} four cases. Bulk magnetization and muon spin relaxation measurements find ground states with long-range magnetic order for the three anhydrous samples, with the latter indicative of similar moment sizes and a 100\% ordered volume fraction. Single crystal neutron diffraction of K$_2$IrBr$_6$ indicates Type I AFM order. While this ordered spin configuration is different than the Type III AFM order previously predicted for (NH$_4$)$_2$IrCl$_6$ \cite{65_harris} and identified for K$_2$IrCl$_6$ \cite{67_hutchings, 68_minkiewicz}, both magnetic structures can be explained by a Heisenberg-Kitaev Hamiltonian that also includes significant NNN Heisenberg exchange. Our comprehensive study demonstrates that the iridium halide family provides an excellent alternative to the more familiar iridates for investigating $J_{\rm eff}$~$=$~1/2 fcc magnetism.

\section{II. Experimental details}

Commercial powders of K$_2$IrCl$_6$, K$_2$IrBr$_6$, (NH$_4$)$_2$IrCl$_6$ and Na$_2$IrCl$_6 \cdotp $ 6(H$_2$O) were purchased from Alfa Aesar.  Single crystals of K$_2$IrCl$_6$, K$_2$IrBr$_6$, and (NH$_4$)$_2$IrCl$_6$ were grown out of a supersaturated water solution, by dissolving the halide salts at $T \sim$ 80$^\circ C$ and slowly cooling to room temperature over a two day period. The largest crystals of K$_2$IrCl$_6$, K$_2$IrBr$_6$ were on the order of 1 - 10 mg and typically had octahedral geometry; the crystals of (NH$_4$)$_2$IrCl$_6$ were similar in shape, but smaller by a factor of two. To investigate the room-temperature crystal structures and assess phase purity, powder x-ray diffraction (XRD) was performed on the commercial powder and crushed-up crystals using a Bruker D8 Advance X-ray Diffractometer with a 1.5406~\AA~wavelength incident beam. Since Na$_2$IrCl$_6 \cdotp $ 6(H$_2$O) can easily be converted to Na$_2$IrCl$_6 \cdotp $ 2(H$_2$O) or Na$_2$IrCl$_6$ depending on the relative humidity level in the air \cite{18_bao}, the crystal structure of this sample was routinely checked by XRD before performing other measurements. If the majority phase was not Na$_2$IrCl$_6 \cdotp $ 6(H$_2$O), an iterative process of soaking the sample in water and then remeasuring XRD was conducted. Neutron powder diffraction (NPD) was additionally performed on polycrystalline K$_2$IrCl$_6$ and K$_2$IrBr$_6$ using the HB-2A powder diffractometer \cite{18_calder} of the High Flux Isotope Reactor (HFIR) at Oak Ridge National Laboratory (ORNL) to investigate the low-temperature crystal structures of these materials. The K$_2$IrBr$_6$ powder was loaded in a vanadium can with a 5~mm inner diameter and the data was collected down to 1.5~K by loading the sample in a cryostat. To ensure that the crystal structure was measured well below the known 3~K magnetic transition, the K$_2$IrCl$_6$ powder was loaded in an aluminum can with the same inner diameter and the data was collected down to 0.3~K, with a He-3 insert that was placed in a helium cryostat. All data was collected with a neutron wavelength of 1.54~\AA~and a collimation of open-21$'$-12$'$.

X-ray absorption spectroscopy (XAS) was performed on polycrystalline samples of K$_2$IrCl$_6$, K$_2$IrBr$_6$, (NH$_4$)$_2$IrCl$_6$ and Na$_2$IrCl$_6 \cdotp $ 6(H$_2$O), as well as the canonical $J_{\rm eff}$~$=$~1/2 magnet Sr$_2$IrO$_4$, at room temperature using the A2 beamline at the Cornell High Energy Synchrotron Source (CHESS) to assess the importance of spin-orbit coupling to the Ir$^{4+}$ electronic ground states. Measurements were collected at both the Ir $L_2$ (2$p_{1/2}$~$\rightarrow$~5$d$) and $L_3$  (2$p_{3/2}$~$\rightarrow$~5$d$) absorption edges, which occur at energies of 12.824~keV and 11.215~keV respectively. The energy of the incident x-ray beam was selected using a diamond-(1 1 1) double crystal monochromator, with higher harmonic contributions suppressed by a combination of Rh-coated mirrors and a 50\% detuning of the second monochromator crystal. The XAS measurements were performed in transmission geometry, using a series of three ion chambers ($I_0$, $I_1$, and $I_2$). The sample was mounted between $I_0$ and $I_1$, while an elemental Ir reference sample was mounted between $I_1$ and $I_2$. This configuration allows a direct measurement of the linear x-ray attenuation coefficient, $\mu(E)$, which is defined by the intensity ratio of the incident and transmitted x-ray beams. In this case, $\mu_{sample}(E)$~$=$~ln($I_0/I_1$) and $\mu_{Ir}(E)$~$=$~ln($I_1/I_2$). The energy calibration of this setup is accurate to within 0.25 eV, and direct comparisons between sample and reference spectra can be used to rule out any systematic energy drifts over the course of the experiment.

Resonant inelastic x-ray scattering (RIXS) measurements were conducted on aligned single crystal samples of K$_2$IrCl$_6$ and K$_2$IrBr$_6$, an unaligned single crystal of (NH$_4$)$_2$IrCl$_6$, and a polycrystalline sample of Na$_2$IrCl$_6 \cdotp $ 6(H$_2$O) using the MERIX spectrometer on beamline 27-ID of the Advanced Photon Source (APS) at Argonne National Laboratory to investigate the Ir$^{4+}$ crystal field excitations. The aligned single crystals were measured at both room temperature and 10~K, while the unaligned single crystal and powder samples were only measured at room temperature. The incident x-ray energy was tuned to the Ir $L_3$ absorption edge at 11.215 keV. A double-bounce diamond-(1 1 1) primary monochromator, a channel-cut Si-(8 4 4) secondary monochromator, and a spherical (2 m radius) diced Si-(8 4 4) analyzer crystal were used to obtain an overall energy resolution of $\sim$~35 meV (full width at half maximum [FWHM]). In order to minimize the elastic background intensity, most of the measurements were carried out in horizontal scattering geometry with a scattering angle close to 90 degrees. For the aligned single crystal samples of K$_2$IrCl$_6$ and K$_2$IrBr$_6$, the actual $Q$ positions measured were $Q$~$=$~(7.26, 7.26, 7.26) and (7.58, 7.58,7.58) rlu respectively. For the bromide system, data was also collected with a scattering angle of 95 degrees corresponding to $Q$~$=$~(7.9, 7.9, 7.9) rlu to confirm lack of momentum dependence for the relevant excitations.  

Bulk thermodynamic data were obtained using instruments housed in the Materials Research Laboratory at the University of Illinois. Magnetization measurements were performed using a Quantum Design MPMS3 SQUID magnetometer and polycrystalline samples in applied fields of $\mu_0 H$ = 5 kOe over a temperature range of $T$ = 2 - 300 K. Heat capacity data were obtained on small unaligned crystals of K$_2$IrCl$_6$, K$_2$IrBr$_6$ and (NH$_4$)$_2$IrCl$_6$  in zero field using a Quantum Design PPMS system equipped with a He-3 insert to allow for measurements down to T = 0.3 K. 

Elastic neutron scattering measurements were performed on the 14.6~meV fixed-incident-energy triple-axis spectrometer HB-1A of the HFIR at ORNL using a K$_2$IrBr$_6$ single crystal on the order of 10~mg. The main goal of this experiment was to determine the magnetic structure for this compound. The experimental background was minimized by using a double-bounce monochromator system, mounting two-highly oriented pyrolytic graphite (PG) filters in the incident beam to remove higher-order wavelength contamination, and placing a PG analyzer crystal before the single He-3 detector for energy discrimination. A collimation of 40$'$-40$'$-40$'$-80$'$ resulted in an energy resolution at the elastic line just over 1 meV (FWHM). The elastic scattering was measured between 4 and 20~K using a closed cycle refrigerator.

Muon spin relaxation ($\mu$SR) measurements were performed at TRIUMF, Canada on polycrystalline samples of K$_2$IrCl$_6$, K$_2$IrBr$_6$ and (NH$_4$)$_2$IrCl$_6$ using the M15 and M20 surface muon beamlines equipped with the Dilution Refrigerator (DR) and LAMPF spectrometers respectively, which have base temperatures of 25~mK and 1.8~K. In a $\mu$SR experiment, positive muons are implanted into the samples one at a time with the stopping sites determined by the locations of local mimina in the Coulomb energy landscape. After implantation, muon spins precess around the local magnetic field at the stopping site, and then each decay into a positron (plus two neutrinos) after a mean lifetime of 2.2 $\mu s$. For a fixed temperature and magnetic field, millions of positron events are recorded by two opposing counters to build up a series of histograms, which can be used to reveal the time-evolution of the muon spin polarization. This quantity is known as the muon asymmetry \cite{97_dereotier, yaouanc_textbook} and is given by: $A(t)$~$=$~$\frac{N_B(t) - \alpha N_F(t)}{N_B(t) + \alpha N_F(t)}$, where $N_B(t)$ and $N_F(t)$ are the number of counts in the two (back and front) positron counters and $\alpha$ is a calibration parameter which was measured using a weak transverse field well above the ordering transitions. Most other measurements were performed in zero-field (ZF) geometry. K$_2$IrBr$_6$ was measured at M20 using LAMPF with an ultra-low background setup, (NH$_4$)$_2$IrCl$_6$ was measured at M15 using the DR where the Ag background from the sample holder is significant, and K$_2$IrCl$_6$ was measured at both beamlines.

\section{III. Crystal structures}

Heavy transition metal halides with the antifluorite structure $A_2 B X_6$ have been studied for a long time \cite{wyckoff_textbook} and are known to frequently crystallize in an ideal fcc structure or lower-symmetry variants. There are intriguing similarities between the antifluorite structure and the commonly-studied double perovskite structure $A_2 B B' X_6$ \cite{15_vasala}, with the former largely resembling the latter with an empty $B'$ site. These vacancies allow the antifluorite structure to accommodate water molecules in a variety of different ways \cite{16_tian, 18_bao}, and one needs to proceed with extreme caution when investigating this material family. Fortunately, the addition of water to the ideal fcc antifluorite structure is seen to lower the crystal symmetry and modify the overall atomic configuration substantially. Thus, x-ray diffraction is capable of detecting water in these compound by probing these symmetry-lowering structural transitions, even  it is not an effective tool for detecting the water molecules directly. With these thoughts in mind, we first use this technique to confirm the crystal structures of the iridium halides K$_2$IrCl$_6$, K$_2$IrBr$_6$, (NH$_4$)$_2$IrCl$_6$ and Na$_2$IrCl$_6 \cdotp $ 6(H$_2$O) at room temperature and assess their phase purity. At the same time, these measurements allow us to establish the local environment of the Ir$^{4+}$ ions, which is an essential first step towards assessing the $J_{\rm eff}$~$=$~1/2 candidacy of these materials.

\begin{figure}
\centering
\scalebox{0.23}{\includegraphics{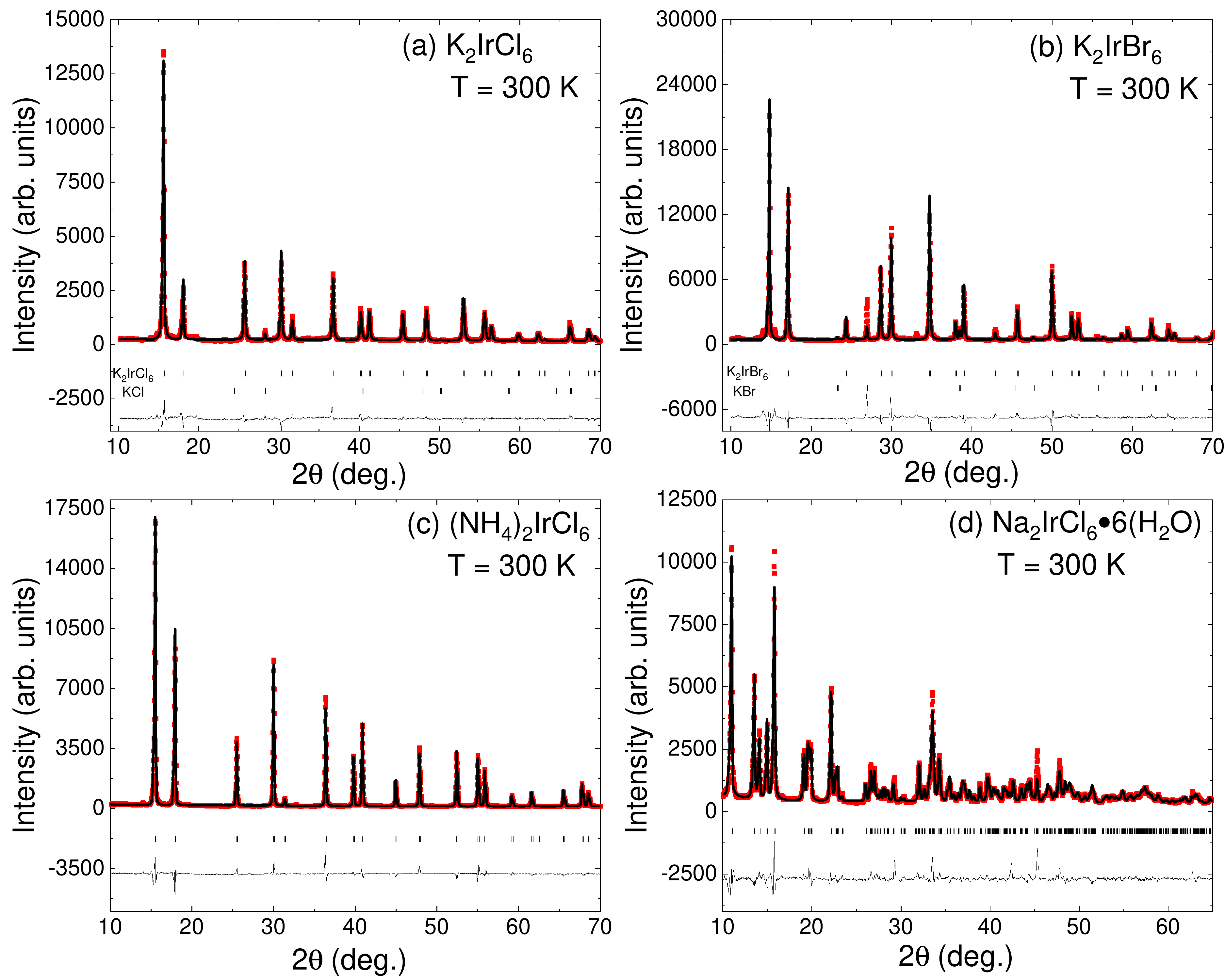}}
\caption{\label{Fig1} (color online) Powder x-ray diffraction data, indicated by the solid symbols and collected with a Cu$_{K\alpha}$ source at room temperature, is shown for (a) K$_2$IrCl$_6$, (b) K$_2$IrBr$_6$, (c) (NH$_4$)$_2$IrCl$_6$ and (d) Na$_2$IrCl$_6 \cdotp $ 6(H$_2$O). The best structural refinements are superimposed on the data as solid curves, the difference curves are shown below the diffraction patterns, and the expected Bragg peak positions are indicated by ticks. Note that a small amount of KCl and KBr impurity were identified in the K$_2$IrCl$_6$ and K$_2$IrBr$_6$ pattern respectively. }
\end{figure}

Our XRD results for these four systems are shown in Fig.~\ref{Fig1}. The data is shown as solid red squares, and the results of Rietveld refinement obtained using FullProf \cite{93_rodriguez} are superimposed on the data as solid black curves. Since XRD is only weakly sensitive to hydrogen atoms, these were excluded from the refinements. Table~\ref{Table1} presents the space group and lattice constants determined by the refinements. We find that K$_2$IrCl$_6$, K$_2$IrBr$_6$, and (NH$_4$)$_2$IrCl$_6$ crystallize in the {\it Fm$\bar{3}$m} space group indicative of the ideal antifluorite structure; this result for K$_2$IrCl$_6$ is consistent with a recently published room-temperature structure determination \cite{19_khan}. The ideal antifluorite structure ensures that the Ir$^{4+}$ ions have cubic point symmetry, which is a required prerequisite for a $J_{\rm eff}$~$=$~1/2 electronic ground state. The higher electronegativity of Cl relative to Br leads to a significantly smaller unit cell for K$_2$IrCl$_6$ and (NH$_4$)$_2$IrCl$_6$ compared to K$_2$IrBr$_6$.

\begin{figure}
\centering
\scalebox{0.23}{\includegraphics{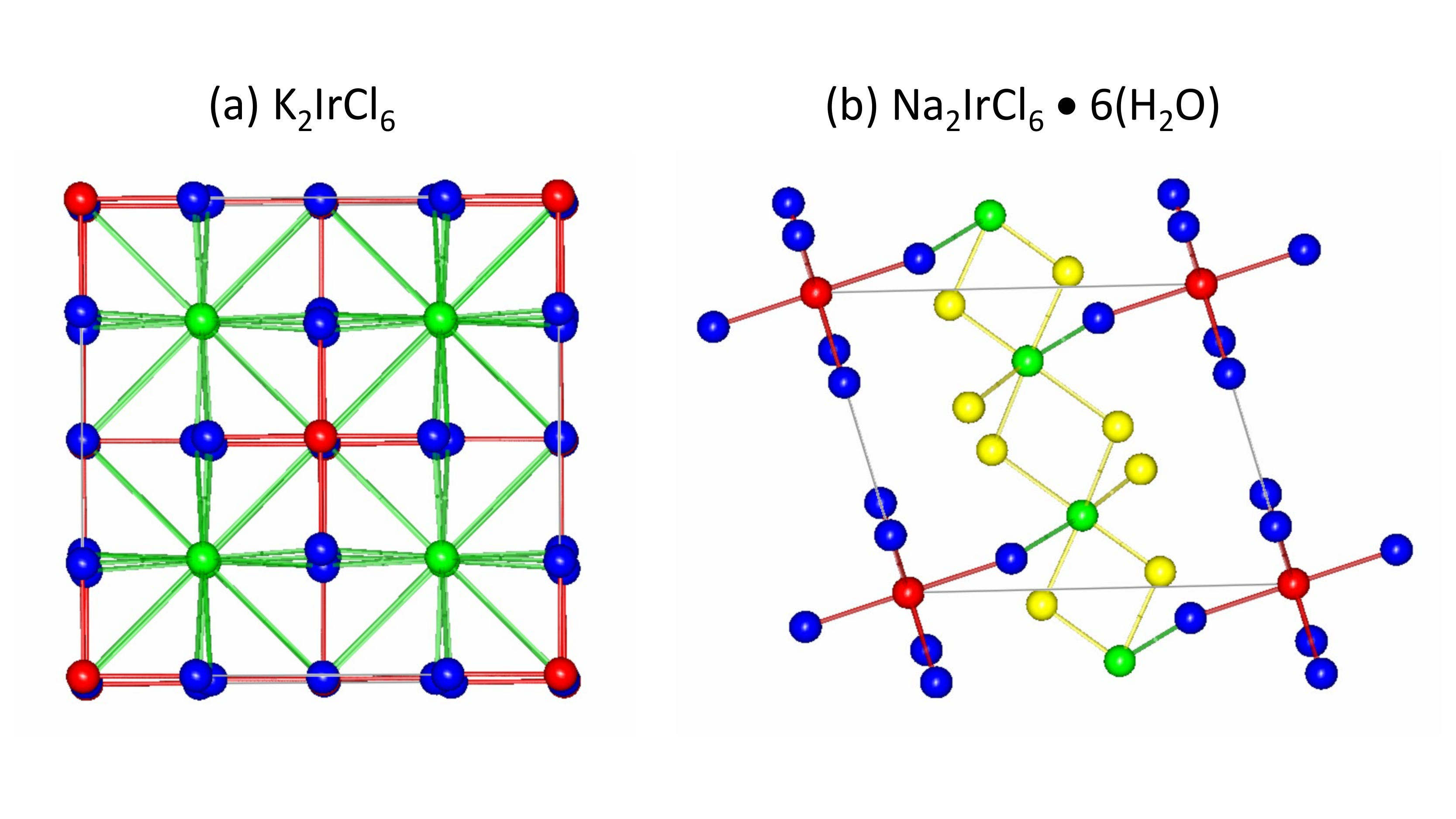}}
\caption{\label{Fig2} (color online) The crystal structure of (a) K$_2$IrCl$_6$ and (b) Na$_2$IrCl$_6 \cdotp $ 6(H$_2$O) viewed along the a-axis of their respective unit cells. The alkali ions, Ir ions, Cl ions, and O ions are shown in green, red, blue, and yellow respectively, while the H ions are omitted to improve clarity. Although the local environment of the Ir ions is nearly identical for the two structure types, the coordination of the alkali ions is drastically different. }
\end{figure}

The Na$_2$IrCl$_6 \cdotp $ 6(H$_2$O) compound is best described by the triclinic space group {\it P$\bar{1}$}, and our XRD refinement result agrees well with previous work \cite{18_bao}. The difference between the ideal antifluorite structure and this lower-symmetry variant is illustrated in Fig.~\ref{Fig2}. In the ideal case, the alkali ions are twelve-fold coordinated with Cl, and the NN distance is quite large ($\sim$~3.45~\AA~for K$_2$IrCl$_6$ at room temperature). This arrangement enables the structure to accommodate water molecules by displacing the alkali ions, which lowers the crystal symmetry. For the cases of Na$_2$IrCl$_6 \cdotp $ 2(H$_2$O) and Na$_2$IrCl$_6 \cdotp $ 6(H$_2$O), the addition of water molecules leads to the formation of NaCl$_4$O$_2$ and NaClO$_5$ octahedra respectively, with significantly shorter NN distances $<$~3~\AA~between the alkali ions and Cl. Notably, the IrCl$_6$ octahedra remain intact with the addition of water molecules to the structure \cite{18_bao}, even as the global symmetry is modified significantly. For Na$_2$IrCl$_6 \cdotp $ 6(H$_2$O), the minimal IrCl$_6$ octahedral distortions suggest that this system may also be close to the ideal $J_{\rm eff}$~$=$~1/2 limit.

\begin{table}[htb]
\begin{center}
\caption{\label{Table1} Space group (SG) and lattice parameters for K$_2$IrCl$_6$, K$_2$IrBr$_6$, (NH$_4$)$_2$IrCl$_6$, and Na$_2$IrCl$_6 \cdotp$ 6(H$_2$O) extracted from Rietveld refinements of room temperature powder XRD data. The lattice constants are in \AA~and all angles are in degrees.}

\begin{tabular}{l l l l l}
\hline
\hline
Material & K$_2$IrCl$_6$ & K$_2$IrBr$_6$ & (NH$_4$)$_2$IrCl$_6$ & Na$_2$IrCl$_6 \cdotp $ 6(H$_2$O) \\
\hline
SG & {\it Fm$\bar{3}$m} & {\it Fm$\bar{3}$m} & {\it Fm$\bar{3}$m} & {\it P$\bar{1}$} \\
$a$ & 9.7720(2) & 10.3120(2) & 9.8663(1) & 6.7339(6) \\
$b$ & 9.7720(2) & 10.3120(2) & 9.8663(1) & 7.0946(6) \\
$c$ & 9.7720(3) & 10.3120(3) & 9.8663(1) & 8.4016(6) \\
$\alpha$ & 90 & 90 & 90  & 102.014(3) \\
$\beta$ & 90 & 90 & 90  & 98.830(3) \\
$\gamma$ & 90 & 90 & 90 & 107.726(3) \\
\hline\hline
\end{tabular}
\end{center}
\end{table}

\begin{figure*}
\centering
\scalebox{0.235}{\includegraphics{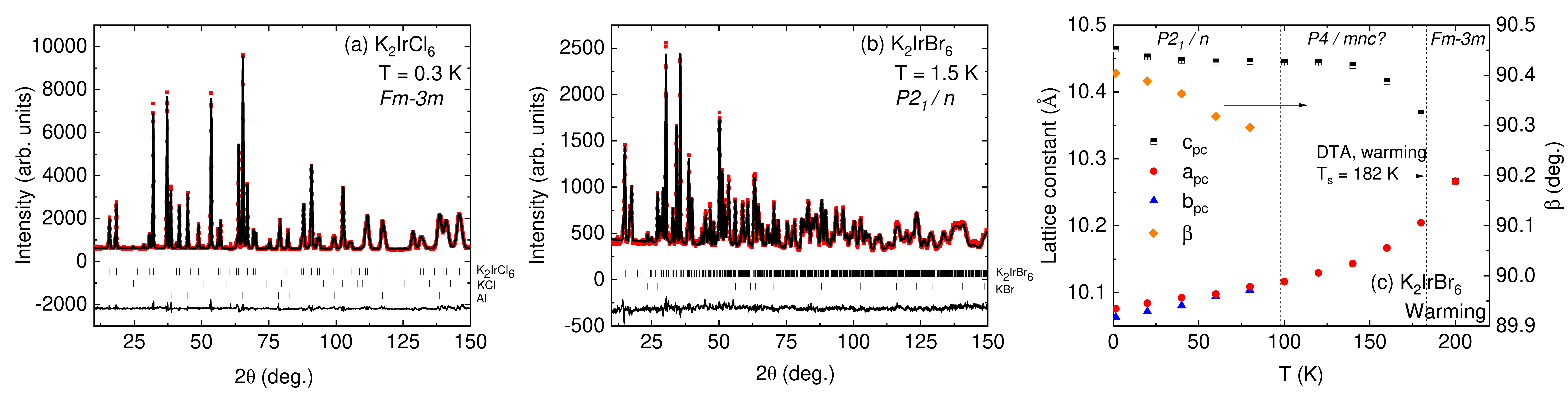}}
\caption{\label{Fig3} (color online) Neutron powder diffraction (NPD) data, indicated by the solid symbols and collected with a neutron wavelength~1.54~\AA, is shown for (a) K$_2$IrCl$_6$ at 0.3~K and (b) K$_2$IrBr$_6$ at 1.5~K. The best structural refinements are superimposed on the data as solid curves, the difference curves are shown below the diffraction patterns, and the expected Bragg peak positions are indicated by ticks. (c) The temperature-dependence of the lattice constants (pseudo-cubic values) for K$_2$IrBr$_6$ extracted from Rietveld refinements of the NPD data collected on warming. Note that the $\sim$180~K onset temperature of the Bragg peak splitting agrees well with the value of $T_s$ obtained from previous differential thermal analysis (DTA) work \cite{77_rossler}. }
\end{figure*}

Polycrystalline samples of the two non-hydrogenous samples were also measured with NPD to look for evidence of structural phase transitions at lower temperatures. While a structural transition for K$_2$IrCl$_6$ at $T_s$~$=$~2.8~K has been proposed previously \cite{79_moses}, synchrotron XRD measurements have only been reported down to 20~K \cite{19_khan}. Temperature-dependent diffraction measurements have not been reported previously for K$_2$IrBr$_6$, although differential thermal analysis reveals two phase transitions of unknown origin at 182~K and 13~K \cite{77_rossler}. Figure~\ref{Fig3}(a) and (b) show our NPD data from the HB2A instrument, plotted as solid red squares. This data was collected using a neutron wavelength of~1.54~\AA~at a temperature of 0.3~K for K$_2$IrCl$_6$ and 1.5~K for K$_2$IrBr$_6$. Rietveld refinement results performed using FullProf \cite{93_rodriguez} are superimposed on the data as black solid curves. No extra peaks indicative of magnetic order are observed in the NPD data below the 3 K magnetic transition for K$_2$IrCl$_6$ or the 13 K transition of unknown origin for K$_2$IrBr$_6$, although in the context of our results given below, we believe this is due to  the extremely small ordered moments.

\begin{table}[htb]
\begin{center}
\caption{\label{Table2}Structural parameters for K$_2$IrCl$_6$ at 0.3~K and K$_2$IrBr$_6$ at both 1.5~K and 200~K extracted from the refinements of the~1.54~\AA~neutron powder diffraction data. The lattice constants and bond distances are in \AA~and all angles are in degrees.}

\begin{tabular}{l l l l l}
\hline
\hline
Material & K$_2$IrCl$_6$ & K$_2$IrBr$_6$ & K$_2$IrBr$_6$& \\
$T$ & 0.3~K & 1.5~K & 200~K \\
\hline
SG & {\it Fm$\bar{3}$m} & {\it P$2_1$/n} & {\it Fm$\bar{3}$m} \\
$a$ & 9.6634(1) & 7.1158(2) & 10.2670(2) \\
$b$ & 9.6634(1) & 7.1248(2) & 10.2670(2) \\
$c$ & 9.6634(1) & 10.4641(2) & 10.2670(2) \\
$\beta$ & 90 & 90.404(2) & 90  \\
K $x$ & 0.25 & 0.509(1) & 0.25 \\
K $y$ & 0.25 & 0.525(1) & 0.25 \\
K $z$ & 0.25 & 0.250(1) & 0.25  \\
Ir & (0, 0, 0) & (0.5, 0, 0.5) & (0, 0, 0)  \\
$X_1$ $x$ & 0.2401(1) & 0.2068(8) & 0.2397(2) \\
$X_1$ $y$ & 0 & 0.2199(9) & 0 \\
$X_1$ $z$ & 0 & 0.9907(4) & 0 \\
$X_2$ $x$ & 0 & 0.2763(9) & 0 \\
$X_2$ $y$ & 0.2401(1) & 0.7129(8) & 0.2397(2) \\
$X_2$ $z$ & 0 & 0.9856(5) & 0 \\
$X_3$ $x$ & 0 & 0.4709(4) & 0 \\
$X_3$ $y$ & 0 & 0.0007(9) & 0 \\
$X_3$ $z$ & 0.2401(1) & 0.2640(2) & 0.2397(2) \\
R$_\mathrm{wp}$ & 4.4\% & 3.9\% & 5.6\% \\
$\chi^2$ & 16.7 & 7.1 & 7.4 \\
Ir-$X_1$ & 2.320(1) & 2.481(6) & 2.461(2) \\
Ir-$X_2$ & 2.320(1) & 2.488(6) & 2.461(2) \\
Ir-$X_3$ & 2.320(1) & 2.477(2) & 2.461(2)  \\
$X_1$-Ir-$X_2$ &  90          &  91.1(3)         & 90        \\
$X_2$-Ir-$X_3$ &  90          &  90.1(3)         & 90        \\
$X_1$-Ir-$X_3$ &  90          &  90.2(3)         & 90         \\
\hline\hline

\end{tabular}
\end{center}
\end{table}

Table~\ref{Table2} shows lattice constants, atomic fractional coordinates, and selected bond distances and angles extracted from the NPD refinements. We find that K$_2$IrCl$_6$ refines well in the space group {\it Fm$\bar{3}$m} corresponding to the ideal antifluorite structure even at 0.3~K, suggesting that there is no structural phase transition at 2.8~K within the resolution of the HB-2A instrument. On the other hand, clear peak splitting is observed in the K$_2$IrBr$_6$ data below $T_s$~$=$~182~K on warming.  Refinements with both the tetragonal {\it P4/mnc} and monoclinic {\it P2$_1$/n} space groups, which are commonly realized in symmetry-lowering transitions of the antifluorite structure, were performed using the 1.5~K base temperature data. Superior agreement factors were found for the monoclinic cell ($R_{wp}$ = 3.9\% and $\chi^2$ = 7.1) as compared to the tetragonal cell ($R_{wp}$ = 7.5\% and $\chi^2$ = 26.1). Similar results were obtained at elevated temperatures, although the size of the monoclinic distortion decreases continuously with increasing temperature. By 100 K, the monoclinic distortion (if present) can no longer be resolved in the HB-2A data and this leads to {\it P2$_1$/n} refinements that do no converge without including unphysical constraints on the atomic positions. Therefore, the data between 100 K and $T_s$ were refined in the tetragonal {\it P4/mnc} space group. Although previous DTA work found no evidence for a second phase transition between 20 K and $T_s$ \cite{77_rossler}, past diffraction work on other antifluorite compounds \cite{78_boysen, 84_abrahams} has revealed two structural transitions following the sequence {\it Fm$\bar{3}$m} $\rightarrow$ {\it P4/mnc} $\rightarrow$ {\it P2$_1$/n}; this behavior may be relevant for K$_2$IrBr$_6$. The temperature-dependence of the lattice parameters extracted from our refinements is shown in Fig.~\ref{Fig3}(c). There is a significant elongation of the c-axis below $T_s$, yet interestingly the Ir-Br bond distances and the Br-Ir-Br bond angles show very small deviations from ideal octahedral geometry in the monoclinic phase; this suggests that the Ir$^{4+}$ electronic ground state may remain close to the ideal $J_{\rm eff}$~$=$~1/2 limit over the entire temperature range investigated.

\section{IV. Single ion properties}

Although it is reasonable to expect that spin-orbit coupling plays a key role in controlling the electronic and magnetic properties of  K$_2$IrCl$_6$, K$_2$IrBr$_6$, (NH$_4$)$_2$IrCl$_6$ and Na$_2$IrCl$_6 \cdotp $ 6(H$_2$O), the strength of the spin-orbit interactions has not been determined in these materials, and in fact has been seen to vary widely in other 5d-compounds \cite{12_clancy}. The strength of spin-orbit interactions can be measured directly through investigation of the branching ratio (BR) with XAS \cite{88_laan, 88_thole_1, 88_thole_2}. The branching ratio is defined as $BR$~$=$~$I_{L_3}/I_{L_2}$, where $I_{L_2}$ and $I_{L_3}$ are the integrated intensities of the white-line features measured at the $L_2$ and $L_3$ absorption edges of Ir, respectively. The branching ratio is directly proportional to the expectation value of the spin-orbit coupling operator $<\vec{L} \cdot \vec{S}>$ \cite{12_clancy}. A branching ratio that is significantly enhanced from the statistical value of 2 is indicative of strong coupling between the local orbital and spin moments, which is a strict requirement for a $J_{\rm eff}$~$=$~1/2 electronic ground state. As a counterexample, the pressure-induced collapse of the $J_{\rm eff}$~$=$~1/2 state in $\alpha$-Li$_2$IrO$_3$ is accompanied by a rapid drop in the BR \cite{18_clancy}, which approaches the value for elemental Ir (BR $\sim$ 3) in its dimerized, non-$J_{\rm eff}$, high pressure state.

\begin{figure*}
\centering
\scalebox{0.9}{\includegraphics{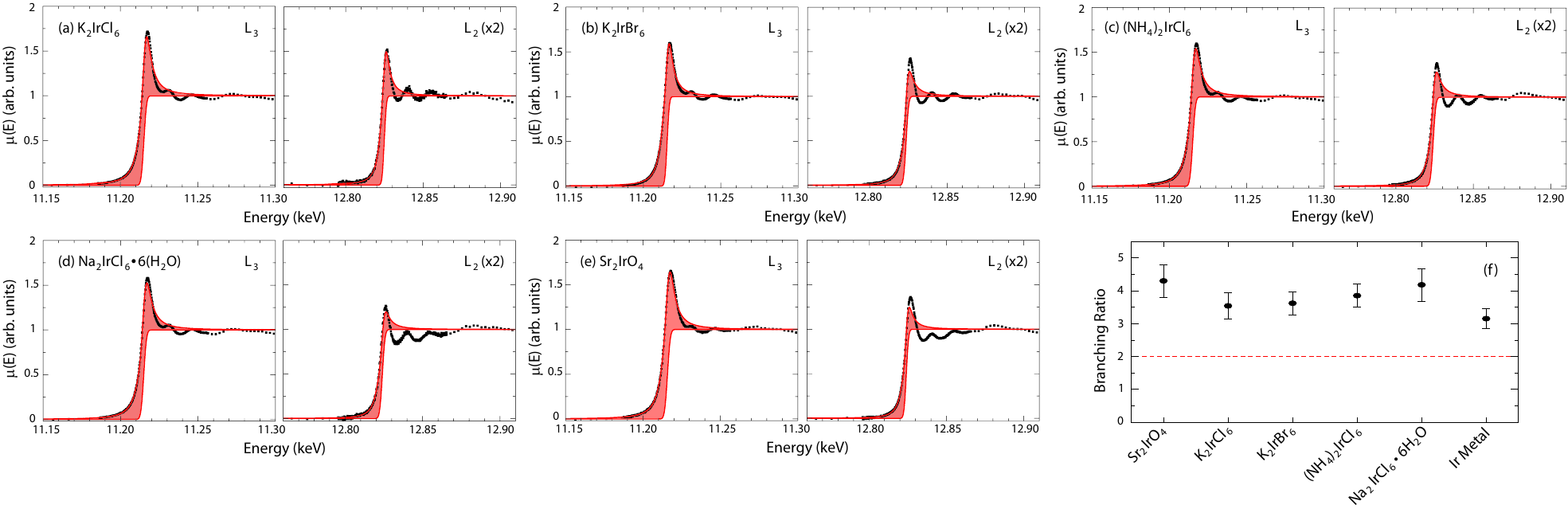}}
\caption{\label{Fig4} (color online) X-ray absorption spectra collected at the Ir $L_3$ edge (left) and Ir $L_2$ edge (right) for (a)  K$_2$IrCl$_6$, (b) K$_2$IrBr$_6$, (c) (NH$_4$)$_2$IrCl$_6$, (d) Na$_2$IrCl$_6 \cdotp $ 6(H$_2$O), and (e) Sr$_2$IrO$_4$. For all samples, the sharp white-line feature is significantly enhanced at the $L_3$ edge as compared to the $L_2$ edge. (f) Experimental branching ratios for K$_2$IrCl$_6$, K$_2$IrBr$_6$, (NH$_4$)$_2$IrCl$_6$, and Na$_2$IrCl$_6 \cdotp $ 6(H$_2$O). The branching ratios for the reference samples Sr$_2$IrO$_4$ and Ir are also shown for comparison purposes.}
\end{figure*}

The x-ray absorption spectra at the Ir $L_3$ and $L_2$ edges for all four iridium halides investigated are plotted in Fig.~\ref{Fig4}(a) - (d) as the linear x-ray attenuation coefficient $\mu(E)$ vs energy, along with data for a Sr$_2$IrO$_4$ reference sample in Fig.~\ref{Fig4}(e). The data were normalized to an edge step of 1 at the $L_3$ edge and 0.5 at the $L_2$ edge, reflecting the ratio of 2p$_{3/2}$ and 2p$_{1/2}$ initial states available for these processes. The background in the pre-edge region of the absorption spectra was modeled by a linear function, while the post-edge region was treated with a quadratic polynomial.  The integrated intensities of the $L_2$ and $L_3$ white-line features, indicated by the shaded regions in Fig.~\ref{Fig4}, were determined using two approaches: (1) by fitting each spectra to a simple Lorentzian (white-line) and arctangent (edge step) fit function, and (2) by numerically integrating the area between the continuum edge step and the experimental data.  A BR was determined for each compound using the average of the white-line intensities obtained from these two approaches. The BRs extracted from this analysis for K$_2$IrCl$_6$, K$_2$IrBr$_6$, (NH$_4$)$_2$IrCl$_6$ and Na$_2$IrCl$_6 \cdotp $ 6(H$_2$O) are shown in Fig.~\ref{Fig4}(f), where they are also compared to the values for Sr$_2$IrO$_4$ and elemental Ir. The statistical BR of two, expected in the limit of negligible SOC, is indicated by the dashed red line.  The enhanced BR values for the four iridium halides are within the range typically found for $J_{\rm eff}$~$=$~1/2 magnets, indicating that the Ir$^{4+}$ ions in these materials may share the same local electronic ground state.

\begin{figure}
\centering
\scalebox{0.9}{\includegraphics{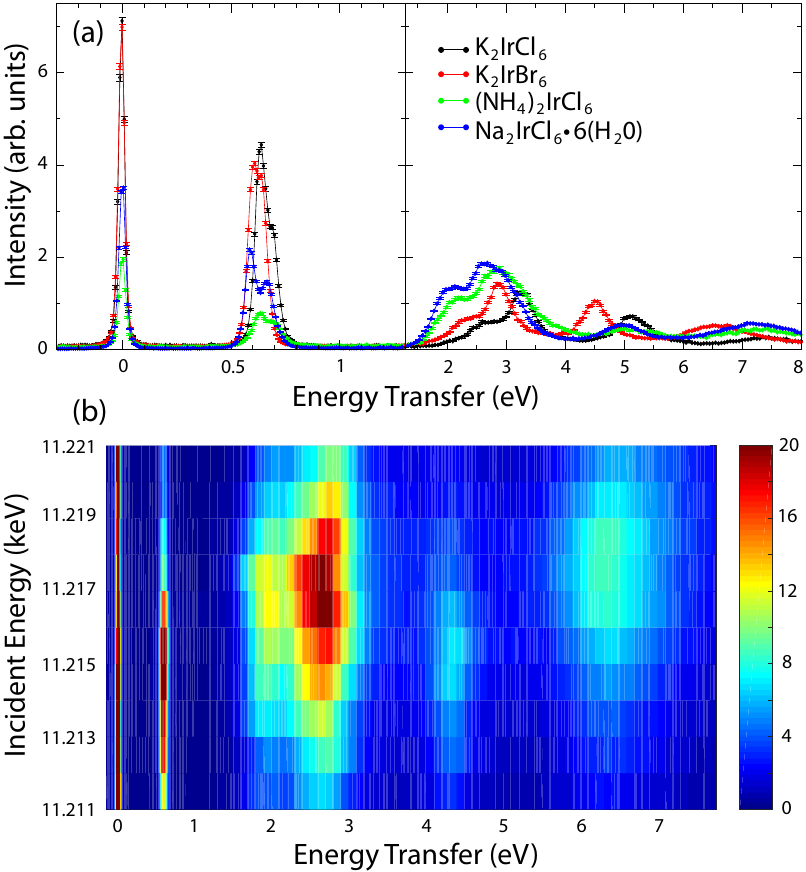}}
\caption{\label{Fig5} (color online) (a) Room temperature resonant inelastic x-ray scattering spectra collected at the Ir $L_3$ edge ($E_i$ = 11.215 keV) for K$_2$IrCl$_6$,  K$_2$IrBr$_6$, (NH$_4$)$_2$IrCl$_6$, and Na$_2$IrCl$_6 \cdotp $ 6(H$_2$O). The K$_2$IrCl$_6$ and K$_2$IrBr$_6$ samples were aligned single crystals, the (NH$_4$)$_2$IrCl$_6$ sample was an unaligned single crystal, and the Na$_2$IrCl$_6 \cdotp $ 6(H$_2$O) sample was polycrystalline. We assign the features between 0.5 - 0.8 eV to intraband $t_{2g}$ excitations, the higher energy features between 1.3 and 3.5 eV to interband $t_{2g}$-$e_g$ excitations, and the highest energy features to charge transfer excitations. (b) Incident energy dependence of the room temperature RIXS spectrum for K$_2$IrBr$_6$. }
\end{figure}

RIXS provides complementary information on the electronic ground state and is a uniquely powerful technique for assessing the proximity of materials to the ideal $J_{\rm eff}$ limit due to its ability to directly probe splittings of the $J_{\rm eff}$~$=$~3/2 quartet excited state arising from non-cubic crystal field terms. Specifically, RIXS measurements of crystal field excitations, in particular the intra-$t_{2g}$ levels, can be used to quantitatively determine the spin-orbit coupling constant $\lambda$ and the non-cubic crystal field $\Delta$ at the Ir$^{4+}$ site. In Fig.~\ref{Fig5}(a), we present the room temperature RIXS spectra at the Ir $L_3$ edge for aligned single crystals of K$_2$IrCl$_6$ and K$_2$IrBr$_6$, an unaligned single crystal of (NH$_4$)$_2$IrCl$_6$, and a polycrystalline sample of Na$_2$IrCl$_6 \cdotp $ 6(H$_2$O). Several peaks are apparent in the data, and based on comparisons with iridates \cite{13_gretarsson, 16_clancy} we assign the features between 0.5 - 0.8 eV to intraband $t_{2g}$ excitations, the higher energy features between 1.3 and 3.5 eV to interband $t_{2g}$-$e_g$ excitations, and the highest energy features to charge transfer excitations.  This assignment is supported by the incident energy dependence of the room temperature RIXS spectrum for K$_2$IrBr$_6$, as illustrated in Fig.~\ref{Fig5}(b). Note that $d-d$ transitions and charge transfer excitations involving $t_{2g}$ excited states resonate at 10{\it Dq} $\sim$ 2 eV below those involving the $e_g$ states.

\begin{figure}
\centering
\scalebox{0.4}{\includegraphics{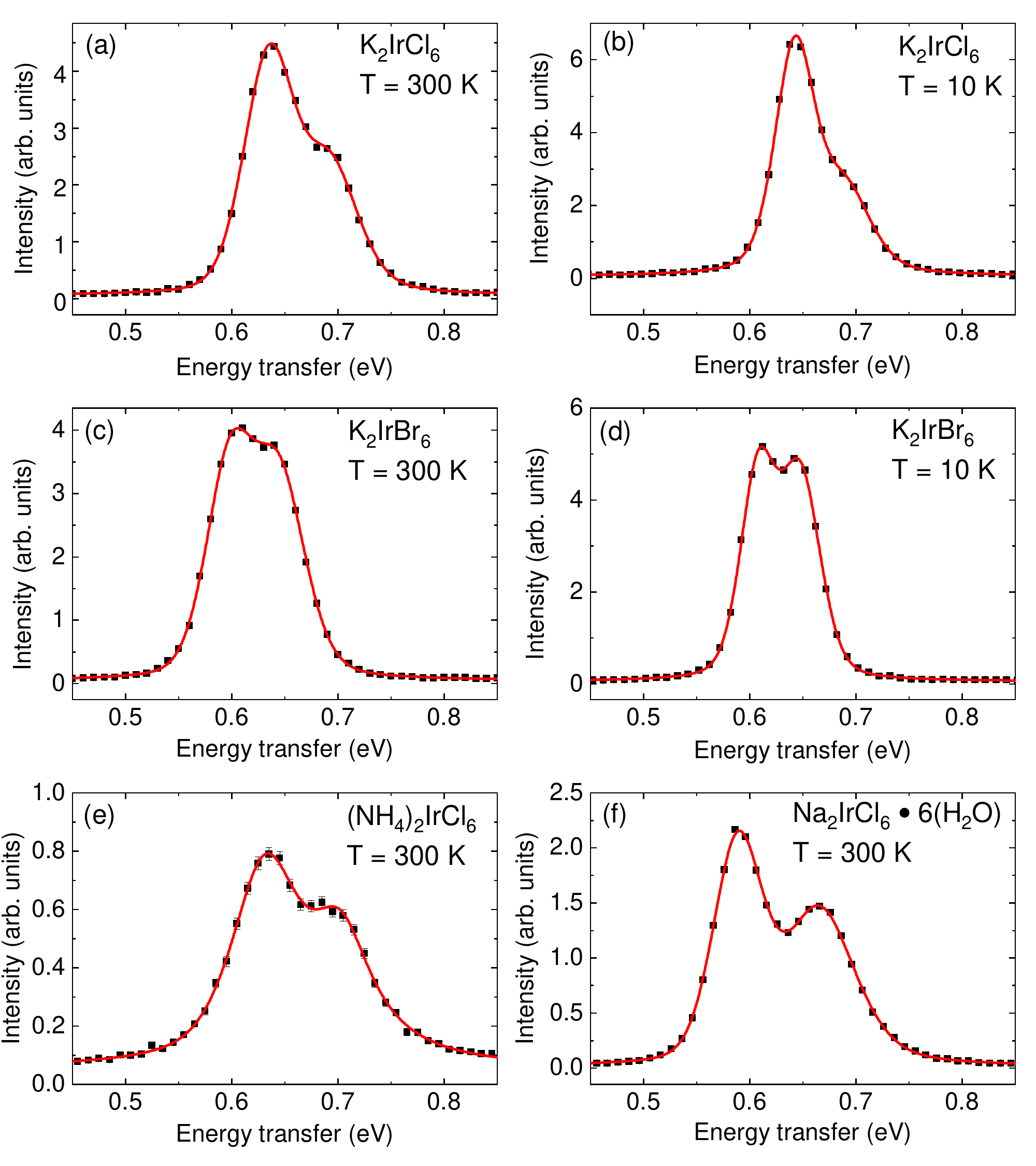}}
\caption{\label{Fig6} (color online) (a) - (f) An enlarged version of the RIXS spectra over a limited energy transfer range for (a) K$_2$IrCl$_6$ at 300 K, (b) K$_2$IrCl$_6$ at 10 K, (c) K$_2$IrBr$_6$ at 300 K, (d) K$_2$IrBr$_6$ at 10 K, (e)  (NH$_4$)$_2$IrCl$_6$ at 300 K, and (f) Na$_2$IrCl$_6 \cdotp $ 6(H$_2$O) at 300 K.  Note the presence of two peaks between energy transfers $\hbar \omega$ of 0.58 to 0.71 eV, even for the cubic samples, corresponding to intraband $t_{2g}$ crystal field transitions. The solid curves represent pseudo-Voigt fitting results. }
\end{figure}

We have plotted an enlarged version of the RIXS spectra in Fig.~\ref{Fig6} to highlight the intra-$t_{2g}$ excitations in the four Ir halides measured in this study. The aligned single crystal samples, K$_2$IrCl$_6$ and K$_2$IrBr$_6$, were also measured at 10~K to check for temperature dependence associated with any structural changes.  The $T$ = 10~K spectra show minimal differences compared to the room temperature spectra.  We confirmed the lack of momentum dependence for these excitations by measuring spectra for K$_2$IrBr$_6$ at both $Q$~$=$~(7.58, 7.58, 7.58) and (7.9, 7.9, 7.9) rlu.  This absence of momentum dependence is typical of $d-d$ excitations in other Ir-based systems with small magnetic bandwidth (e.g. the honeycomb, pyrochlore, and double perovskite iridates).  The $Q$~$=$~(7.58, 7.58, 7.58) spectra are plotted in Fig.~\ref{Fig6}(c) and (d).  It is immediately obvious that {\it all} data sets show a small peak splitting indicative of deviations from an ideal $J_{\rm eff}$~$=$~1/2 state, even for samples where diffraction measurements revealed cubic symmetry. Thus, RIXS results for the cubic samples are reminiscent of previous work on the fcc double perovskite Ba$_2$CeIrO$_6$ \cite{19_aczel, 19_revelli}, where the intra-$t_{2g}$ peak splitting was attributed to local tetragonal distortions. A similar explanation may be applicable for the cubic samples studied here, although we cannot rule out extremely small distortions of the global structures or second order dynamic Jahn-Teller distortions.  It is also interesting to note that the intra-$t_{2g}$ peak splitting in K$_2$IrBr$_6$ does not increase below the structural phase transition at $T_s$~$=$~182~K.  In fact, the splitting appears to slightly decrease.  A similar decrease is observed in the low temperature peak splitting of K$_2$IrCl$_6$, in spite of the absence of an analogous structural transition.

To extract quantitative information from this data, we fit each RIXS spectrum to the sum of two pseudo-Voigt functions representing the two intraband $t_{2g}$ transitions. These fits were used to establish precise peak positions ($\hbar \omega_1$ and $\hbar \omega_2$) and enable a meaningful quantitative comparison between the various samples. The fitted results are presented in Table~\ref{Table3}. As discussed in Ref.~\cite{19_aczel}, we attribute the energy difference between the $\hbar \omega_1$ and $\hbar \omega_2$ peaks to the non-cubic crystal field splitting $\Delta$, while the average energy of these two peaks is $\frac{3}{2} \lambda$. This method provides a reliable estimate of $\lambda$ when $\Delta$~$\le$~200~meV \cite{14_sala_2}.

\begin{table}[htb]
\begin{center}
\caption{\label{Table3} RIXS fitting results of the intra-$t_{2g}$ excitations, spin-orbit coupling constants ($\lambda$), and non-cubic crystal field splitting ($\Delta$) of the $J_{\rm eff}$~$=$~3/2 manifold for selected iridium halides and iridates. All parameters, except for dimensionless $\Delta/\lambda$, are in meV.}
\begin{tabular}{l l l l l l l}
\hline
\hline
Material & $\hbar \omega_1$ & $\hbar \omega_2$ & $\lambda$ & $\Delta$ & $\Delta/\lambda$ & Ref. \\
\hline
Sr$_3$Ir$_2$O$_7$ & 500 & 700 &  400  &  200  & 0.5 & \cite{17_lu}   \\
Sr$_2$IrO$_4$ & 550 & 700 & 417 & 150 & 0.36 & \cite{17_lu} \\
Sr$_2$CeIrO$_6$ & 645(1) & 760(3) & 468(1) & 115(3) & 0.25 & \cite{19_aczel} \\
Ba$_2$CeIrO$_6$ & 625(1) & 735(4) & 453(1) & 110(4) & 0.24 & \cite{19_aczel} \\
K$_2$IrF$_6$ & 802(1) & 914(1) & 572(1) & 112(1) & 0.20 & \cite{17_rossi} \\
Rb$_2$IrF$_6$ & 805(1) & 915(1) & 573(1) & 110(1) & 0.19 & \cite{17_rossi} \\
Na$_2$IrCl$_6 \cdotp $ 6 H$_2$O & 589(1) & 665(1) & 418(1) & 76(1) & 0.18 & this work \\
(NH$_4$)$_2$IrCl$_6$ & 631(1) & 702(2) & 444(1) & 71(2) & 0.16 & this work \\
K$_2$IrCl$_6$ (300 K) & 635(1) & 693(2) & 443(1) & 58(2) & 0.13 & this work \\
K$_2$IrCl$_6$ (10 K) & 642(1) & 690(2) & 444(1) & 48(2) & 0.11 & this work \\
K$_2$IrBr$_6$ (300 K) & 598(1) & 645(1) & 414(1) & 47(1) & 0.11 & this work \\
K$_2$IrBr$_6$ (10 K) & 609(1) & 652(1) & 420(1) & 43(1) & 0.10 & this work \\
\hline\hline
\end{tabular}
\end{center}
\end{table}

Table~\ref{Table3} compares several key parameters for iridates and iridium halides extracted from the analysis of RIXS data, including the spin-orbit coupling constant $\lambda$ and the non-cubic crystal field splitting $\Delta$. The materials are sorted according to decreasing values of the $\Delta/\lambda$ ratio, as this is arguably the best measure of a material's proximity to the ideal $J_{\rm eff}$~$=$~1/2 state. Some trends are immediately apparent upon careful inspection of this table. As expected, materials with connected IrO$_6$ octahedra (i.e. Sr$_3$Ir$_2$O$_7$ and Sr$_2$IrO$_4$) have wider $d$-bands and therefore yield comparatively large values of $\Delta/\lambda$ relative to the other materials in the table which feature isolated Ir$X_6$ octahedra. For this subclass of materials, it appears that the value of $\Delta/\lambda$ is best minimized by incorporating an anion with a low electronegativity into the crystal structure. This feature leads to larger lattice constants, decreased energy scales for the intra-$t_{2g}$ crystal field excitations, larger Ir-Ir distances, and ultimately narrower $d$-bands. Therefore, K$_2$IrBr$_6$ is the closest to the ideal $J_{\rm eff}$~$=$~1/2 limit, followed by the chloride samples. Local structure modifications of the Ir$X_6$ octahedra or temperature changes appear to have significantly less impact on the value of $\Delta/\lambda$, although small decreases in this parameter are observed as the octahedra become more ideal or the temperature is lowered. These results indicate that iridium halides incorporating anions from the third or fourth row of the periodic table (or beyond) offer unprecedented proximity to the desirable $J_{\rm eff}$~$=$~1/2 electronic ground state.

\begin{figure*}
\centering
\scalebox{0.37}{\includegraphics{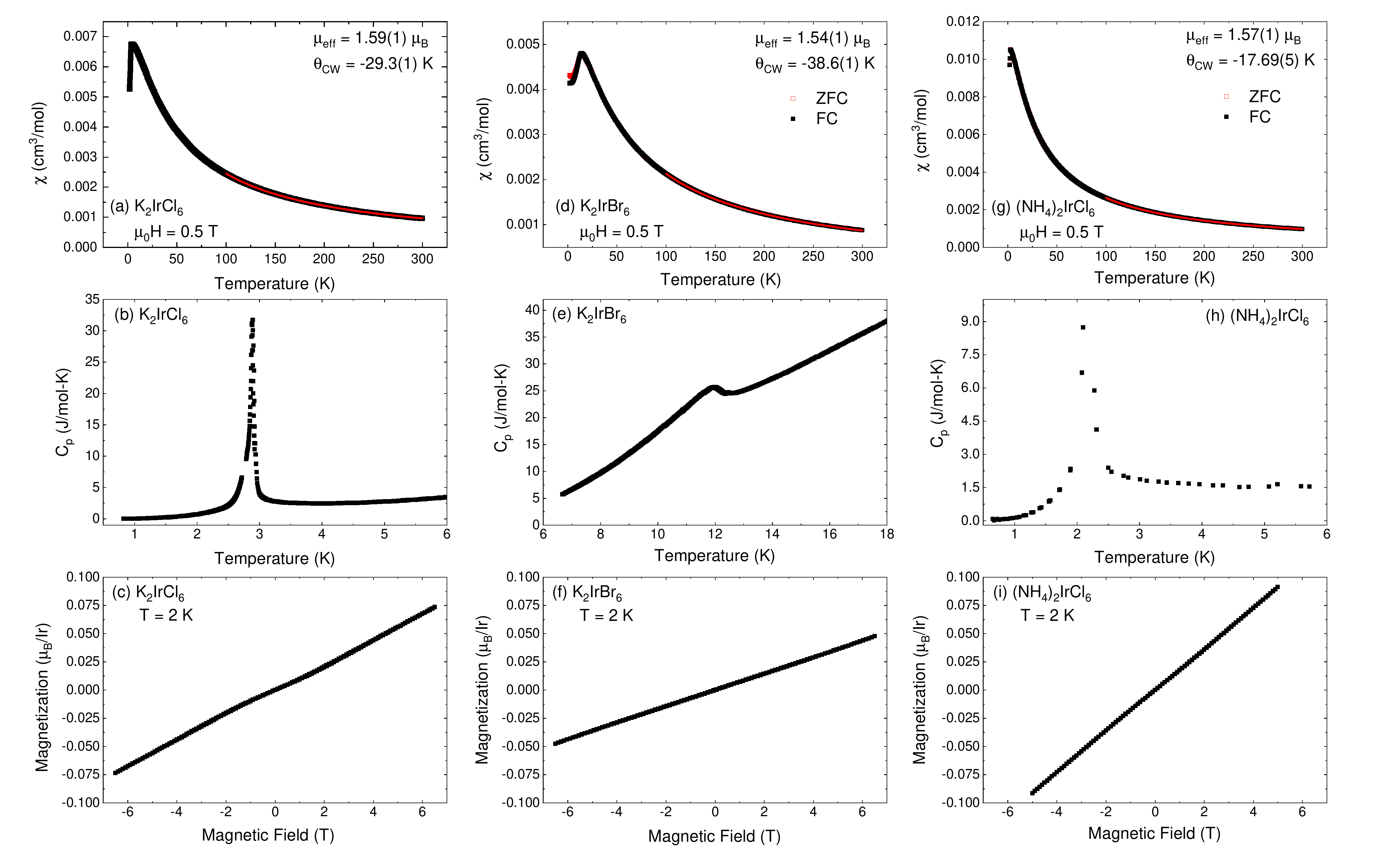}}
\caption{\label{Fig7} (color online) Magnetic susceptibility $\chi$ vs temperature, heat capacity $C_p$ vs temperature, and magnetization $M$ vs field are shown for: (a) - (c) K$_2$IrCl$_6$, (d) - (f) K$_2$IrBr$_6$, and (g) - (i) (NH$_4$)$_2$IrCl$_6$. The $\chi$ and $M$ data were measured using polycrystalline samples, while the $C_p$ data was collected using small, unaligned crystals.}
\end{figure*}

\section{V. Magnetic Ground States}

\subsection{A. Bulk Characterization}

We now focus on the magnetic ground states of the $J_{\rm eff}$~$=$~1/2 systems K$_2$IrCl$_6$, K$_2$IrBr$_6$, and (NH$_4$)$_2$IrCl$_6$. Na$_2$IrCl$_6 \cdotp $ 6 (H$_2$O) is excluded from this section because recent work has shown that it remains paramagnetic down to 1.8~K \cite{18_bao}, and our preliminary characterization is largely in line with these results. Figure~\ref{Fig7} presents the magnetic susceptibility $\chi$ (plotted as $M/H$) vs temperature with $\mu_0 H$~$=$~0.5~T and the magnetization $M$ vs field at $T$~$=$~2~K for polycrystalline samples of the three compositions. The heat capacity $C_p$ vs temperature is also shown in the same figure, taken in zero field with small, unaligned crystals.

The magnetic susceptibility above 100~K for each sample was fit to the Curie-Weiss law to extract effective moments and Weiss temperatures, which are also presented in Fig.~\ref{Fig7}. The effective moment values are close to the expectation of 1.73~$\mu_B$ for a $J_{\rm eff}$~$=$~1/2 electronic ground state (assuming $g$~$=$~2) and the negative Weiss temperatures are indicative of dominant antiferromagnetic exchange. The sharp drop in $\chi$ combined with the coincident anomalies in the $C_p$ data arise from antiferromagnetic transitions at $T_N$~$\sim$~2.9~K for K$_2$IrCl$_6$, 12~K for K$_2$IrBr$_6$, and 2.2~K for NH$_4$IrCl$_6$; these values are in excellent agreement with previous work \cite{59_bailey, 59_cook}. The $C_p$ anomalies for K$_2$IrCl$_6$ and (NH$_4$)$_2$IrCl$_6$ are quite sharp and symmetric as expected for a first-order phase transition, while the $C_p$ anomaly for K$_2$IrBr$_6$ has the characteristic $\lambda$ shape for a second order transition. We also note that the linear $M$ vs $H$ results are consistent with all three systems exhibiting collinear antiferromagnetic order below their respective $T_N$ values.


\subsection{B. Magnetic Structure of K$_2$IrBr$_6$}

Previous neutron diffraction work on K$_2$IrCl$_6$ has identified a Type-III AFM structure \cite{67_hutchings, 68_minkiewicz}, which is expected for an fcc Heisenberg magnet with an antiferromagnetic NN $J_1$ and an antiferromagnetic NNN $J_2$ with a value between 0 and $\frac{1}{2}J_1$ \cite{63_lines, 01_lefmann}. The same spin configuration has also been predicted for (NH$_4$)$_2$IrCl$_6$ by measuring $J_1$ and $J_2$ values with electron spin resonance \cite{65_harris} to determine its location on this phase diagram. More recent theoretical work using both classical and quantum treatments has explored how strong spin-orbit coupling modifies the Heisenberg phase diagram by adding anisotropic terms to the Hamiltonian including NN Kitaev and/or off-diagonal exchange interactions \cite{15_cook, 19_revelli, 19_khan}. One important finding of these latter works is that NN anisotropy does not provide an alternative way to stabilize a Type-III AFM structure; NNN interactions therefore cannot be neglected in the iridium halide family, although $J_2$ is expected to be quite weak compared to $J_1$. It is also intriguing that the addition of an AFM Kitaev interaction to the $J_1$-$J_2$ model, which has already been discussed in the context of the double perovskite iridates \cite{15_cook, 16_aczel, 19_aczel, 19_revelli} and K$_2$IrCl$_6$ \cite{19_khan}, leads to a rich phase diagram with several possible collinear ordered states and more exotic incommensurate and spin liquid states \cite{19_revelli} in the classical and quantum cases, respectively.

\begin{figure}
\centering
\scalebox{0.17}{\includegraphics{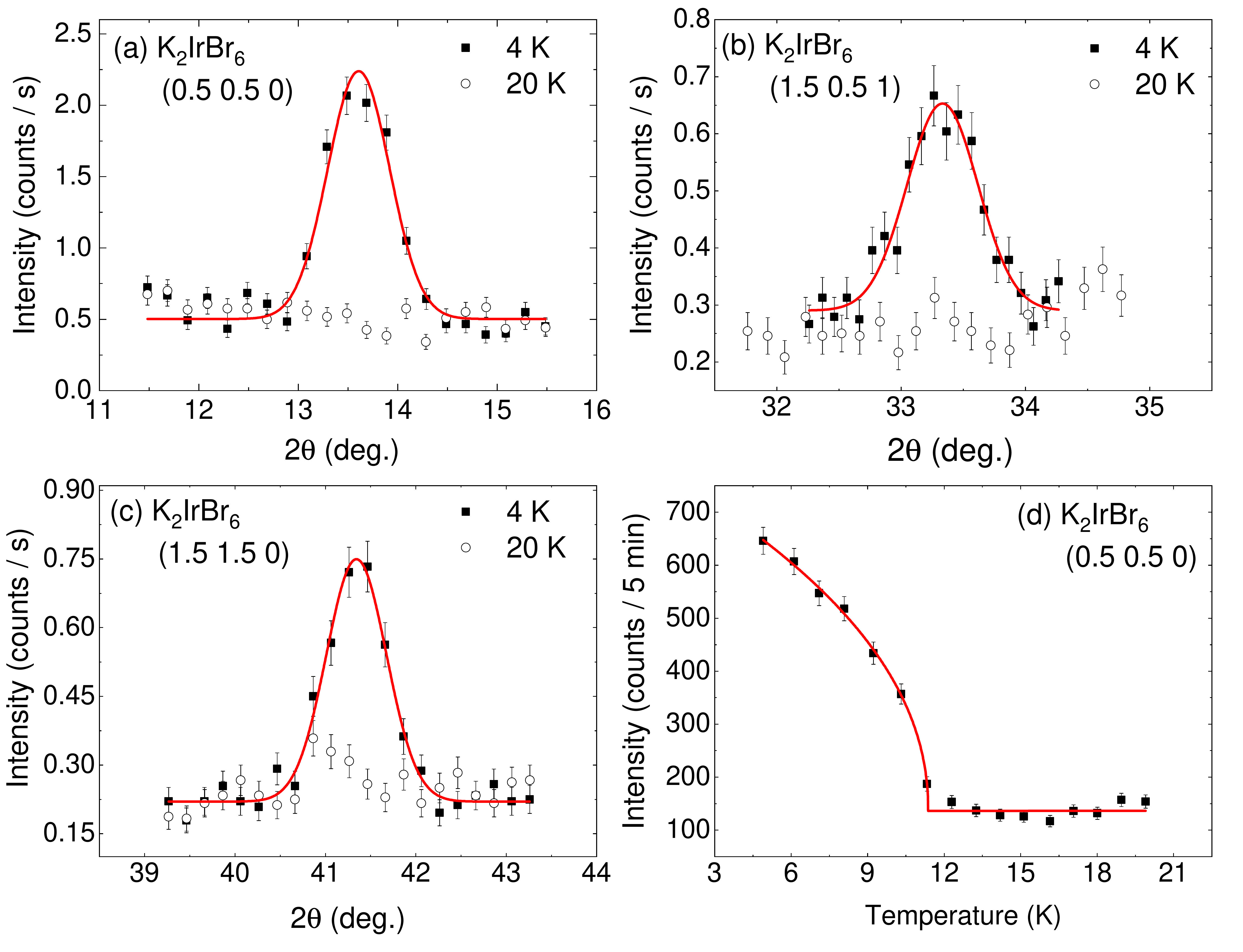}}
\caption{\label{Fig8} (color online) $\theta$-2$\theta$ scans at 4 and 20 K for the (a) (0.5 0.5 0), (b) (1.5 0.5 1), and (c) (1.5 1.5 0) positions in reciprocal space. Bragg peaks develop in each case below $T_N$, which shows that they have a magnetic origin. (d) Intensity vs temperature for the (0.5 0.5 0) magnetic Bragg peak. A power law fit is superimposed on the data and yields $T_N$~$=$~11.37(5)~K. }
\end{figure}

Despite the potential for unconventional magnetism on the fcc lattice, the magnetic ground state of K$_2$IrBr$_6$ has not been predicted or determined until now. We performed elastic neutron scattering measurements on our largest single crystal of K$_2$IrBr$_6$ (on the order of 10 mg) using the HB-1A spectrometer to address this issue. We aligned the sample in the monoclinic (H K H-K)$_m$ scattering plane for this experiment to ensure access to Bragg peak positions associated with many common quasi-fcc magnetic propagation vectors, including $\vec{k_m}$~$=$~0, (0.5 0.5 0), and (0.5 0 0.5), which correspond to Type I and Type II AFM ordered states \cite{09_makowski}. We could not access Type III AFM Bragg peaks in this experiment geometry since $\vec{k_m}$~$=$~(0.5 0.5 0.5) in this case. As shown in Fig.~\ref{Fig8}, we identified several Bragg peaks consistent with a $\vec{k_m}$~$=$~(0.5 0.5 0) propagation vector. The order parameter plot for the (0.5 0.5 0) peak is illustrated in Fig.~\ref{Fig8}(d) and a power law fit reveals $T_N$~$=$~11.37(5)~K, which is in excellent agreement with the magnetic transition temperature established from the bulk characterization measurements described above. Therefore, our neutron results indicate that K$_2$IrBr$_6$ hosts Type I AFM order with the FM planes stacked along the pseudo-cubic [100] direction. Unfortunately, due to the limited magnetic Bragg peaks measured and the formation of structural domains below $T_s$, we were not able to determine a moment size or direction from this data.

The evolution from Type III AFM order in K$_2$IrCl$_6$ to Type I AFM order in K$_2$IrBr$_6$ cannot arise simply from a reduced AFM $J_2$ due to the larger unit cell, so we conclude that the magnetic structure change is instead driven by the monoclinic structural distortion. It is intriguing that both systems appear to be close to a magnetic phase boundary and therefore their magnetic ground states can likely be tuned by external stimuli quite easily. In fact, there is preliminary evidence for a field-induced magnetic structure change between 5 and 6 T in the case of K$_2$IrCl$_6$ \cite{01_meschke}. Future single crystal neutron diffraction studies of both materials in a magnetic field are highly desirable.

\subsection{C. Muon Spin Relaxation}

\begin{figure*}
\centering
\scalebox{0.37}{\includegraphics{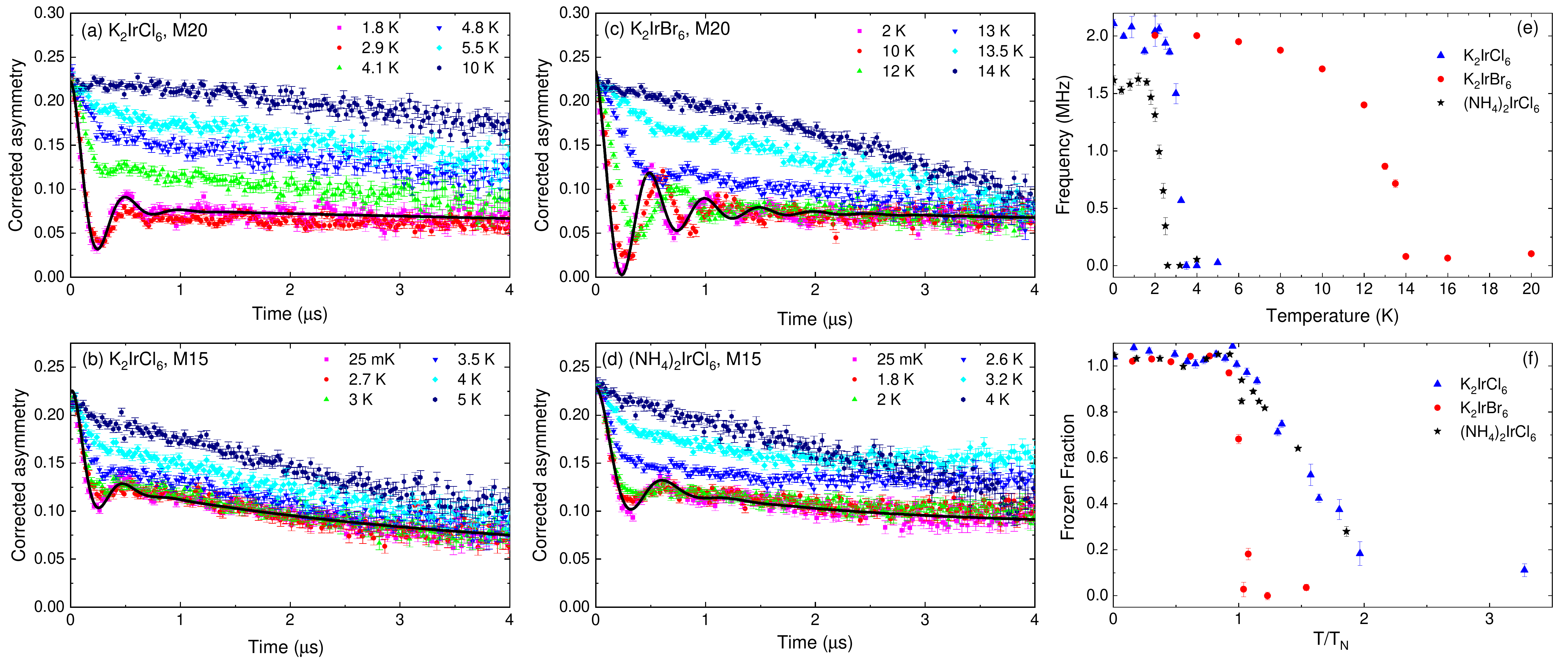}}
\caption{\label{Fig9} (color online) Zero-field $\mu$SR asymmetry spectra vs time at selected temperatures for polycrystalline (a) K$_2$IrCl$_6$ (M20), (b) K$_2$IrCl$_6$ (M15), (c) K$_2$IrBr$_6$ (M20), and (d) (NH$_4$)$_2$IrCl$_6$ (M15). The fit using the functional form described in the main text is superimposed on the base-temperature data in each case. (e) Zero field frequency vs temperature for the three samples. (f) Frozen volume fractions vs temperature for the three samples. }
\end{figure*}

Finally, we also performed muon spin relaxation ($\mu$SR) on polycrystalline samples of K$_2$IrCl$_6$, K$_2$IrBr$_6$, and (NH$_4$)$_2$IrCl$_6$ to assess the homogeneity of the magnetic ground states and to compare the ordered moment sizes in the three compounds. $\mu$SR is a real-space, local probe of magnetism, and therefore it is an excellent method for measuring magnetic volume fractions. The extreme sensitivity of the muon spin to local fields also facilitates the detection of extremely weak magnetic moments that cannot be identified by neutron diffraction due to the signal-to-noise limitations arising from neutron absorption or other complexities of this technique, such as the formation of the poorly-resolved structural domains below $T_s$ in K$_2$IrBr$_6$. Notably, the measured frequency in a zero field $\mu$SR experiment is directly proportional to the ordered moment in the material. The time-evolution of the zero-field muon spin polarization, plotted as the muon asymmetry \cite{97_dereotier, yaouanc_textbook}, is depicted in Fig.~\ref{Fig9}(a) - (d) for K$_2$IrCl$_6$, K$_2$IrBr$_6$, and (NH$_4$)$_2$IrCl$_6$ at various temperatures. We observe a heavily-damped oscillatory signal for all three samples that is characteristic of long-range magnetic order with appreciable decoherence (short $1/T_2$).

The signature of homogeneous magnetic order (i.e. 100\% ordered volume fraction) for a polycrystalline sample in a zero field $\mu$SR experiment is a signal that is well-described by the following function:
\begin{equation}
A(t) = \frac{2A_{tot}}{3} e^{-t/T_{2}}\cos(\omega t) + \frac{A_{tot}}{3} e^{-t/T_1},
\end{equation}
where $1/T_2$ is the dephasing or decoherence rate, $1/T_1$ is the spin-lattice relaxation rate, and $A_{tot}$ is the total sample asymmetry determined from a weak transverse field measurement. The non-oscillating term (the `1/3 tail') represents the fraction of the initial muon spins which are parallel to the local field at the muon site, which reduces the precession amplitude. If the ordered fraction was to decrease from 100\%, one would expect the non-oscillating component to increase from 1/3 at the expense of the oscillating component.

Our $\mu$SR data was taken using two separate instruments at the TRIUMF facility. Powders of K$_2$IrCl$_6$ and K$_2$IrBr$_6$ were first measured on the M20 beamline using a helium flow cryostat and an ultra low-background sample holder, which ensures that nearly all of the measured signal comes from muons which have landed in the sample of interest. We initially fit these data at low temperatures to a modified version of Eq.~(1), where the amplitude of the oscillating and non-oscillating terms was allowed to vary independently. In this way, we confirmed the expected 2:1 ratio for all temperatures $T < T_N$, where the Neel temperature is that determined by heat capacity measurements. This is consistent with an ordered system involving the full sample volume. A second set of measurements were performed on powders of K$_2$IrCl$_6$ and (NH$_4$)$_2$IrCl$_6$ using a dilution refrigerator on the M15 beamline to access temperatures as low as 25 mK, well below the ordering transitions for the two materials. These data were fit to the same function as above, plus a weakly-relaxing background term $A_{bkg}e^{-\lambda t}$ to account for muons which landed in the silver sample holder. For these fits, the 2:1 ratio of oscillating and non-oscillating components from the sample was assumed for the base temperature fits to determine the amplitude of the background component; the near-identical background amplitude for both materials lead us to extend our conclusion of homogenous magnetic order to (NH$_4$)$_2$IrCl$_6$, as determined above  for K$_2$IrCl$_6$ and K$_2$IrBr$_6$. For subsequent fits, the background amplitude was fixed and the amplitudes of the sample components were released, but were found to evolve in a way which maintained the same 2:1 ratio for all temperatures $T< T_N$.

Representative data from each of the four sets of measurements are plotted in panels Fig.~\ref{Fig9}(a) - (d), where curves of best fit for the lowest temperatures are superimposed on the data. The extracted frequencies at base temperatures are 2.11(4), 2.01(1), and 1.62(4)~MHz for K$_2$IrCl$_6$, K$_2$IrBr$_6$, and (NH$_4$)$_2$IrCl$_6$, respectively. The nearly identical frequencies for the K$_2$IrCl$_6$ and K$_2$IrBr$_6$ data is somewhat surprising given the different lattice and spin structures in the two compounds, as described above, but might be an indication that the implanted muons are binding to the apical halogen anions in the nearly ideal Ir$X_6$ octahedra. In this scenario, the data implies that the ordered moment in K$_2$IrBr$_6$ is comparable to the value of $m_{ord} = 0.80~\mu_B$ previously determined for K$_2$IrCl$_6$ \cite{76_lynn}.  On the other hand, (NH$_4$)$_2$IrCl$_6$ and K$_2$IrCl$_6$ share the same ideal cubic antifluorite structure and, it has been argued \cite{65_harris}, the same Type III AFM ground state. The data for these two materials are therefore more directly comparable, and the reduced frequency for (NH$_4$)$_2$IrCl$_6$ likely indicates that the ordered moment in this compound is reduced by $\sim$~25\% as compared to K$_2$IrCl$_6$.

In Fig.~\ref{Fig9}(e), we plot the fitted frequencies as a function of temperature for the three materials. The temperature evolution is typical for magnetic materials below their ordering transitions, and frequencies obtained for K$_2$IrCl$_6$ from the two sets of measurements agree in the temperature range where they overlap. Approaching the respective transitions, the frequencies drop sharply with increasing temperature for K$_2$IrCl$_6$ and (NH$_4$)$_2$IrCl$_6$, while the decrease is much more gradual for K$_2$IrBr$_6$. These observations are consistent with our specific heat results presented above, which indicated first order magnetic phase transitions for K$_2$IrCl$_6$ and (NH$_4$)$_2$IrCl$_6$ and a second order magnetic phase transition for K$_2$IrBr$_6$.

At temperatures $T > T_N$, the data for all the three samples feature a crossover regime that is not well-described by Eq.~(1) or the non-oscillatory, single component relaxation function typically expected above magnetic ordering transitions in $\mu$SR. Instead, spectra in this region are characterized by a heavily damped oscillatory/spin-glass-like component which gradually decreases in amplitude with increasing temperature, and eventually evolves into the expected weakly-relaxing signal at high temperatures. This intermediate region might be associated with a critical slowing down of short-ranged correlations; similar behavior has been observed in other quantum magnets \cite{17_ziat}. To quantify this crossover regime, we fit the slowly-relaxing part of the spectra ($t>2$ $\mu$s) for all three materials to the simple power exponential function $A_\mathrm{ZF} = A_s(1 - 2f/3)e^{(-t/T_1)^\beta}$, where $A_s$ is the asymmetry of the slowly-relaxing component and $f$ represents the ordered volume fraction \cite{20_chen}. The temperature-dependence of $f$ for these samples is shown in Fig.~\ref{Fig9}(f), with the temperature scale normalized to the transition temperatures determined above. The crossover regime for each sample extends from $T_N$ to the lowest temperature where the paramagnetic volume fraction reaches 100\%, and is significantly larger in the fcc magnets K$_2$IrCl$_6$ and (NH$_4$)$_2$IrCl$_6$ than in the monoclinic compound K$_2$IrBr$_6$. This is likely due to the greater degree of frustration in the cubic materials.

\section{VI. Conclusions}
In this work, we have used x-ray techniques and neutron diffraction to identify a family of iridium halides [i.e. K$_2$IrCl$_6$, K$_2$IrBr$_6$, (NH$_4$)$_2$IrCl$_6$, and Na$_2$IrCl$_6 \cdotp $ 6(H$_2$O)] consisting of Ir$^{4+}$ ions with unprecedented proximity to the desirable $J_{\rm eff}$~$=$~1/2 electronic ground state. We argue that the nearly-ideal $J_{\rm eff}$~$=$~1/2 magnetism arises from a combination of the large spatial separation of the Ir ions and the lower electronegativity of the anions in these compounds, as compared to the well-studied iridates. We also explored the low-temperature magnetic properties of these materials, and find homogeneous magnetic order in the three anhydrous materials with similar moment sizes. We used single crystal neutron diffraction to determine a Type-I AFM spin configuration for K$_2$IrBr$_6$, in contrast to the known Type-III AFM ordered ground state in K$_2$IrCl$_6$. Both the Type-I and Type-III AFM ordered ground states of the anhydrous materials can be explained by a Heisenberg-Kitaev model that also includes NNN Heisenberg interactions \cite{19_revelli}. Future work should aim to quantify the strength of the Kitaev interactions $K$ in iridium halides with the antifluorite structure and to explore the additional phase space afforded by these materials in an effort to identify the elusive spin liquid state that appears on the $J_1$-$J_2$-$K$ quantum phase diagram for fcc magnets.

\section{acknowledgments}
A portion of this research used resources at the High Flux Isotope Reactor, which is a DOE Office of Science User Facilities operated by Oak Ridge National Laboratory. This research used resources of the Advanced Photon Source, a US Department of Energy (DOE) Office of Science User Facility operated for the DOE Office of Science by Argonne National Laboratory under contract No. DE-AC02-06CH11357. Research conducted at CHESS is supported by the National Science Foundation via Awards DMR-1332208 and DMR-1829070. Synthesis, crystal growth, powder X-ray, magnetization and heat capacity measurements were carried out in part in the Materials Research Laboratory Central Research Facilities, University of Illinois. Work by G.J.M., D.R.-i-P., T.A.J., K.L., Q.C. and H.D.Z. was supported by the National Science Foundation, Division of Materials Research under awards DMR-1455264 and DMR-2003117.

\end{document}